\documentclass[aps,pre,floats,twocolumn,showpacs,superscriptaddress]{revtex4}

\usepackage{epsfig}
\usepackage{latexsym,amsmath}

\begin{document}
\title{Clustering in complex networks. II. Percolation properties}

\author{M. \'Angeles Serrano}

\affiliation{School of Informatics, Indiana University,\\ Eigenmann
Hall, 1900 East Tenth Street, Bloomington, IN 47406, USA}

\author{Mari{\'a}n Bogu{\~n}{\'a}}

\affiliation{Departament de F{\'\i}sica Fonamental, Universitat de
  Barcelona,\\ Mart\'{\i} i Franqu\`es 1, 08028 Barcelona, Spain}

\date{\today}

\begin{abstract}
The percolation properties of clustered networks are analyzed in detail. In the case of weak clustering, we present an analytical approach that allows to find the critical threshold and the size of the giant component. Numerical simulations confirm the accuracy of our results. In more general terms, we show that weak clustering hinders the onset of the giant component whereas strong clustering favors its appearance. This is a direct consequence of the differences in the $k$-core structure of the networks, which are found to be totally different depending on the level of clustering. An empirical analysis of a real social network confirms our predictions.
\end{abstract}

\pacs{89.75.-k,  87.23.Ge, 05.70.Ln}

\maketitle

\section{Introduction}
\label{percolation}

The general framework introduced in the preceding paper~\cite{preceding} to
analyze clustering in complex networks provide us with the necessary
tools to tackle the study of the percolation properties of clustered
networks. The introduction of clustering in percolation analysis
represents a theoretical challenge due to the fact that previous
analytical approaches, applied to random two-point correlated
directed and undirected
networks~\cite{Molloy:1995,Molloy:1998,Havlin:2000,Callaway:2000,Newman:2001b,Havlin:2002,Newman:2002a,Vazquez:2003,Boguna:2005},
are based on the idea of branching process. One starts from a given
vertex and follows all the edges attached to it. Then, the process
is repeated again starting from the neighboring vertices avoiding
the source one. All the vertices reachable from a given one belong
to the same connected component and the network is said to be in the
percolated phase when the largest component reaches proportions of
the order of the network size, becoming the so-called giant
connected component (GCC). This scheme works well when the network
is locally tree-like and, thus, the clustering coefficient is very
small. Real networks, however, are shown to have a significant level
of clustering --even for very large networks-- that may change the
percolation properties significantly.

Aside from the theoretical importance of percolation itself, there
is a significant practical interest. Percolation is strongly related
to epidemic processes. In fact, the simplest model for epidemic
spreading, the Susceptible-Infected-Removed
model~\cite{Kermack:1927,May:1991}, can easily be mapped into a bond
percolation
problem~\cite{Grassberger:1983,Sander:2002,Sander:2003,Newman:2002b}.
In its simplest formulation, an infected individual becomes infected
for a random time $t_r$ following a Poisson process, that is,
$\psi_r (t_r)=\delta e^{-\delta t_r}$. On the other hand, when an
infected individual is in contact with a susceptible one, it takes a
random time $t_i$ to infect it, this process being a Poisson process
as well, that is, $\phi_i(t_i)=\lambda e^{-\lambda t_i}$. The
probability that an infected individual infects a susceptible
neighbor before it becomes removed is then
\begin{equation}
p_{inf}=\int_0^{\infty} \phi_i (t) \Psi_r(t) dt
=\frac{\lambda}{\lambda+\delta},
\end{equation}
where $ \Psi_r(t)=\int_t^{\infty} \psi_r(\tau) d\tau$ is the
probability that the vertex remains infected for a time larger than
$t$. Since the infection uses the network as a template to spread,
the infection process can be understood as a bond percolation
problem over the original network when each edge is removed with
probability $q_{inf}=1-p_{inf}$. The percolation threshold
corresponds in this mapping to the onset of pandemic infections
whereas the size of the giant connected cluster corresponds to the
number of infected individuals.

In this second paper, we present analytical results for percolation
in random networks with weak transitivity, that is, with
degree-dependent clustering coefficient, $\bar{c}(k)$, below $(k-1)^{-1}$, which
extends and completes material previously published in
\cite{Serrano:2006}. In the case of percolation in the presence of
strong transitivity --with $\bar{c}(k)>(k-1)^{-1}$ in a given domain--, we present here new counterintuitive results
which demonstrate that, in the percolation process, strong
clustering favors the onset of the giant component whereas weak
clustering hinders it. Furthermore, we show how these outcomes
explain previous results by other authors~\cite{Shi:2006,Newman:2003c}. We also discuss that
these properties are intimately related to the structural
organization of networks, in particular to their k-core
decomposition~\cite{Seidman:1983,Dorogovtsev:2006}. We end this paper by taking a look at the
Pretty-Good-Privacy (PGP) web of trust~\cite{Boguna:2004b}, a large
social network which turns out to be a nice example of a real system
where our predictions apply.

\section{Percolation in weakly clustered networks}

The standard percolation formalism based on branching processes
overcounts the size of a node's second neighborhood when clustering is
present. To correct for this effect, the usual procedure can be
modified in the following way. One starts from a given vertex and
follows all its edges. However, once placed in one of the neighbors,
the next edges to follow are those not pointing to the neighborhood
of the source vertex~\cite{Newman:2003d} (the edges pointing to the neighborhood of the
source vertex are the ones responsible for clustering). It is worth
to notice that, even in this scheme, we are neglecting the fact that
higher order loops may be present in the network. In particular,
squares will connect vertices in the first neighborhood with a
common vertex in the second neighborhood, over-counting it.
Besides, when the multiplicity of the edges is large --by multiplicity, $m$, we mean the number of triangles passing through an edge--, squares
induced by the merge of two triangles that share an edge appear.
Therefore, the implementation of the strategy outlined above will
become exact in the case of disjoint triangles but it will represent an upper
bound to the real size of the giant component when triangles jam into edges. As explained in the preceding paper~\cite{preceding}, weakly clustered networks are those which have a degree-dependent clustering coefficient $\bar{c}(k)<(k-1)^{-1}, \forall k$, implying that edge multiplicity is small and triangles disjoint. In contrast,  strongly clustered networks have high edge multiplicity and, thus, triangles are forced to share edges.

With this in mind, let us start by defining the probability that a
given vertex has $s$ reachable vertices, $G(s)$. For very
heterogeneous networks this function is not very representative and
we have to define the same probability but conditioned to the degree
of the source vertex, $G(s|k)$. These two functions are trivially
related through
\begin{equation}
G(s)=\sum_k P(k) G(s|k)
\end{equation}
where $P(k)$ is the degree distribution. Finally, we need to
introduce an extra function, $g(s|k)$, which measures the
probability that a vertex can reach $s$ other vertices given that it
is connected to a vertex $v$ of degree $k$, and that it cannot visit
either $v$ nor its neighborhood. This last condition guaranties that
we do not over-count contributions due to triangles. We can find a
recursion relation for this function taking into account that now
the process is a branching one with the constraint that at each
generation point we can only use the free edges to continue the
process. Then, function $g(s|k)$ satisfies
\begin{widetext}
\begin{equation}
g(s|k)=\sum_{k'} \sum_m P(k'|k) \phi(m|kk')
\sum_{s_1,s_2,\cdots}g(s_1|k') g(s_2|k') \cdots g(s_{k'-m-1}|k')
\delta_{s,1+s_1+s_2+\cdots s_{k'-m-1}},
\label{eq:g(s|k)}
\end{equation}
\end{widetext}
where $P(k'|k)$ is the probability that a vertex of degree $k$ is connected to a vertex of degree $k'$\footnote{Remember that the two-vertex probability distribution, $P(k,k')$, is related to this transition probability by means of
\[ P(k,k')=\frac{kP(k)}{\langle k \rangle} P(k'|k)
\]} and where we have defined $\phi(m|kk')$ as the probability that an edge
connecting two vertices of degrees $k$ and $k'$ has multiplicity
$m$. The multiplicity matrix $m_{kk'}$ can be computed as the first moment of $\phi(m|kk')$, {\it i. e.},
\begin{equation}
m_{kk'}=\sum_{m=0}^{m^c_{kk'}} m\phi(m|kk'),
\end{equation}
where $m^c_{kk'}=min(k,k')-1$. For randomly assembled networks, we can make use of the
probabilistic interpretation of the edge clustering coefficient $\bar{c}(k,k')=m_{kk'}/m^c_{kk'}$~\cite{Radicchi:2004,preceding} and write that
\begin{equation}
\phi(m|kk')=\left(
\begin{array}{c}
m^c_{kk'}\\
m
\end{array}
\right) [\bar{c}(k,k')]^m[1-\bar{c}(k,k')]^{m^c_{kk'}-m}.
\label{eq:phi(m)}
\end{equation}
In principle, this particular form of the distribution of $m$ is to
be taken as an approximation. However, as we will see, in the case
of randomly assembled networks it works extremely well.

We define the generating function of $g(s|k)$ as
\begin{equation}
\hat{g}(z|k)\equiv \sum_s z^s g(s|k).
\end{equation}
Using this transformation, Eq.~(\ref{eq:g(s|k)}) reads
\begin{equation}
\hat{g}(z|k)=z \sum_{k'} \sum_m P(k'|k) \phi(m|kk')
\left[\hat{g}(z|k')\right]^{k'-m-1}, \label{eq:fundamental}
\end{equation}
which is a closed set of equations for the functions $g$'s. Finally,
$G(s|k)$ and $g(s|k)$ are related through
\begin{equation}
\hat{G}(z|k)=z \left[ \hat{g} (z|k)\right]^k.
\end{equation}
The percolation transition takes place when
Eq.~(\ref{eq:fundamental}), evaluated at $z=1$, admits as a stable
solution $\hat{g}(z=1|k)=\xi (k)\le 1$, that is, there is a finite
probability ($1-\xi(k)$) that the branching process extends up to
infinity, meaning that a giant connected component has been formed. Since $\hat{g}(z=1|k)=1$ is always a fixed point of
Eq.~(\ref{eq:fundamental}), the onset of the giant component is the
point at which this solution becomes unstable. To perform the
stability analysis, we linearize Eq.~(\ref{eq:fundamental}) by
plugging in a solution of the form $\hat{g}(z=1|k)\approx
1+\chi(k)\epsilon$. In the limit $\epsilon \rightarrow 0$, this
operation yields
\begin{equation}
\chi(k)=\sum_{k'} (k'-1-m_{kk'}) P(k'|k)  \chi(k')
\end{equation}
The critical percolation point is ruled by the maximum eigenvalue,
$\Lambda_m$, of the matrix $(k'-1-m_{kk'}) P(k'|k)$. When
$\Lambda_m>1$, the solution $\hat{g}(z=1|k)=1$ becomes unstable
leading the network to the percolated phase, in which a macroscopic
fraction of the system becomes globally connected. The relative size
of the GCC can then be computed as
\begin{equation}
gcc=1-\langle [\xi(k)]^k \rangle.
\end{equation}

\begin{figure}[t]
\epsfig{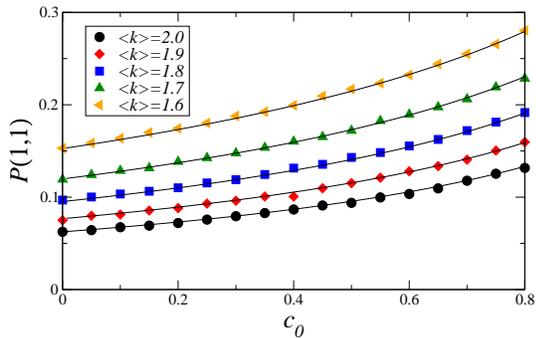}
 \caption{Values of the parameter $P(1,1)$ as a function of $c_0$
obtained from numerical simulations as compared to the analytical
solution given by Eq.~(\ref{eq:P(1,1)b}), for different values of the
average degree and an exponential degree distribution. Symbols are numerical simulations using the algorithm of Ref.~\cite{Serrano:2005b} and solid lines correspond to the analytical solution.} \label{p11}
\end{figure}
\begin{figure}[t]
\epsfig{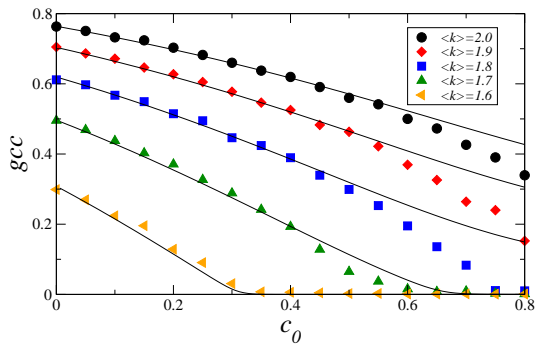}
 \caption{Relative size of the giant component in the case of
 $\bar{c}(k)=c_0/(k-1)$ as a function of $c_0$ for different values
 of the average degree. Symbols are numerical simulations using the
 algorithm of Ref.~\cite{Serrano:2005b} for an exponential degree
 distribution. Solid lines correspond to the numerical solution of
 the set of equations (\ref{eq:fundamental}).} \label{giant1}
\end{figure}

To be able to derive an analytic expression for the percolation
condition, we have to assume some simplifications at this point. On
one hand, we take $m_{kk'}=m_0$, with $m_0$ a constant within the
interval $[0,1]$, that is, we restrict to networks with weak
transitivity. In this situation, making use of the closure condition derived in the preceding paper,
\begin{equation}
\sum_{k'}m_{kk'}P(k,k')=k(k-1)\bar{c}(k)\frac{P(k)}{\langle k \rangle},
\label{cl}
\end{equation}
we can see that the degree-dependent clustering
coefficient takes the form $\bar{c}(k)=c_0(m_0) (k-1)^{-1}$, where
$c_0(m_0)$ is a certain function of $m_0$ to be determined. The second simplification assumes the absence
of two-vertex correlations but, when clustering is present,
uncorrelation can be attained for all degree classes except for the
degree class $k=1$ (see discussion in the preceding paper~\cite{preceding}). The
reason is that vertices of degree $k=1$ cannot participate in
triangles and, therefore, connections involving these vertices are
forced to follow a different pattern. Consequently, assuming the factorization of $P(k,k')$ for $k,k'>1$, the transition probability $P(k'|k)$ takes the form
\begin{equation}
P(k'|k)=\left\{
\begin{array}{lr}
\displaystyle{ \frac{1-2\frac{P(1)}{\langle k
\rangle}+P(1,1)}{(1-\frac{P(1)}{\langle k \rangle})^2}}
\frac{k'P(k')}{\langle k \rangle} & k\wedge k' > 1\\[0.5cm]
\displaystyle{\frac{1-\frac{P(1,1)}{P(1)}\langle k
\rangle}{(1-\frac{P(1)}{\langle k \rangle})}} \frac{k'P(k')}{\langle
k \rangle} &
\begin{array}{r}
k=1\wedge k' > 1\\
k>1\wedge k'=1
\end{array}\\[0.5cm]
\displaystyle{\frac{P(1,1)}{P(1)}}\langle k \rangle & k=k'=1
\end{array}
\right.
\label{eq:P(k'|k)}
\end{equation}
where $P(1,1)$ is the probability that a randomly chosen edge
connects two nodes of degree 1. If clustering is absent, then
$m_0=0$ and $P(1,1)$ can factorize as $P(1,1)=P^2(1)/\langle k
\rangle^2$. Otherwise, if clustering is finite, we can assume that
all edges that are not involved into triangles are realized in the
most random form, which leads to the following expression
\begin{equation}
P(1,1)=\frac{P^2(1)}{\langle k \rangle^2} \frac{1}{1-E_{c}/E},
\label{eq:P(1,1)}
\end{equation}
where $E_c$ is the number of edges participating in triangles and
$E$ is the total number of edges. $E_c$ can be computed as
\begin{equation}
E_c=\frac{1}{2}\sum_{i,j} a_{ij} \Theta(m_{ij}-1),
\end{equation}
where $\Theta(m_{ij}-1)$ is the Heaviside step function. This
equation can now be rewritten as
\begin{equation}
E_c=E \sum_{k,k'=2} P(k,k') \langle \Theta(m-1) |k,k' \rangle,
\end{equation}
where the last average is taken over the set of edges connecting nodes of
degrees $k$ and $k'$, that is,
\begin{equation}
\langle \Theta(m-1) |k,k' \rangle=\sum_{m=1} \phi(m|kk')=1-\phi(0|kk'),
\end{equation}
which can be computed using Eq.~(\ref{eq:phi(m)}). Using this
expression, combined with the two-point correlation function,
Eq.~(\ref{eq:P(k'|k)}) and Eq.~(\ref{eq:P(1,1)}), we can write the
following equation for $P(1,1)\equiv x$,
\begin{equation}
x^2-\left(\frac{\langle \phi \rangle'}{1-\langle \phi
\rangle'}+\frac{2P(1)}{\langle k
\rangle}\right)x+\frac{P^2(1)}{\langle k \rangle^2 (1-\langle \phi
\rangle')}=0, \label{eq:P(1,1)b}
\end{equation}
where $\langle \phi \rangle'$ is the average of $\phi(0|kk')$ over
the set of vertices of degrees larger than 1. $P(1,1)$ corresponds
to the smallest solution of this quadratic equation. Finally, using Eq.~(\ref{cl}), the
clustering factor $c_0(m_0)$ can be written as
\begin{equation}
c_0(m_0)=m_0\frac{1-2\frac{P(1)}{\langle k
\rangle}+P(1,1)}{(1-\frac{P(1)}{\langle k \rangle})}.
\label{c_0}
\end{equation}
Using the results above, the maximum eigenvalue of the matrix
$(k'-1-m_{kk'}) P(k'|k)$ can be analytically computed, yielding a
percolation condition given by
\begin{equation}
\frac{\langle k(k-1)\rangle}{\langle k \rangle} >(1+c_0(m_0))
\frac{m_0}{c_0(m_0)} (1-\frac{P(1)}{\langle k \rangle}).
\label{critical}
\end{equation}
It is easy to prove that the right hand side of this equation is
always larger or equal to $1$. This means that {\it weakly clustered
networks percolate at a higher density of connections as compared to
the unclustered ones}. For very low clustering, this term converges
to 1 and so we recover the percolation threshold of random networks
with a given degree distribution.

There is a particular case in which the symbiosis of a specific form
of the degree distribution and a weak clustering in the frontier,
$\bar{c}(k)=(k-1)^{-1}$, maintains the critical point unchanged with
respect to the unclustered classical random graph. This model is
studied in detail in Ref.~\cite{Shi:2006} as a natural extension of
the Erd\"os-R\'enyi random graph~\cite{Erdos:1959,Erdos:1960}, where
each possible triangle among a fixed number of vertices is realized
with a given probability. In the thermodynamic limit, and for sparse
networks, all edges in the ensemble have fixed multiplicity, $m=1$, and the
generating function of the resulting degree distribution is
$\hat{P}(z)=\exp({\langle k \rangle (z^2-1)/2})$. Since odd degrees
are not present, $P(1)=P(1,1)=0$ and the critical condition
Eq.~(\ref{critical}) becomes $\langle k(k-1)\rangle/\langle k
\rangle>2$ which, for this particular degree distribution translates
into $\langle k \rangle >1$, that is, the same percolation condition
that applies for the classical random graph.

To check the accuracy of the formalism, we generate clustered random
networks with the algorithm introduced in~\cite{Serrano:2005b} with
an exponential degree distribution and a degree-dependent clustering
coefficient of the form $\bar{c}(k)=c_0 (k-1)^{-1}$\footnote{To
generate uncorrelated networks, we set $\beta=1$ in the algorithm of Ref.~\cite{Serrano:2005b}, which guaranties the factorization of $Q(k,k')$~\cite{preceding} and, hence, the lack of degree-degree correlations other than the ones corresponding to degree 1.}. Notice that, when generating the networks, we fix the value of $c_0$ and not $m_0$, as it is done in the derivation above. However, we can make use of Eq.~(\ref{c_0}) to go from one parameter to the other.
Fig.~\ref{p11} shows a perfect agreement between the empirical
values for the parameter $P(1,1)$ as a function of $c_0$ for
different values of the average degree and the analytical results
derived from Eq.~(\ref{eq:P(1,1)b}). In Fig.~\ref{giant1}, we
compare the relative size of the giant connected component as a
function of $c_0$ with the numerical solution of
Eqs.~(\ref{eq:fundamental}) combined with the transition probability
given in Eq.~(\ref{eq:P(k'|k)}). As it can be seen in the figure,
the effect of clustering is to reduce the size of the giant
connected component. The effect is so strong that, in networks with
a moderate average degree, it completely fragments the network when
$c_0$ exceeds a critical value. In other cases, the reduction of the
size can be of more of the fifty percent. For values of $c_0 \in
[0,0.5]$, the agreement between our theory and numerical simulations
is excellent. Beyond this point, our approximation slightly
overestimates the GCC's size. This is mainly due to the fact that in
this regime, links of multiplicity larger than 1 appear which, in
turn, induces the presence of some loops of order four.

\section{Percolation in strongly clustered networks}
\begin{figure}[t]
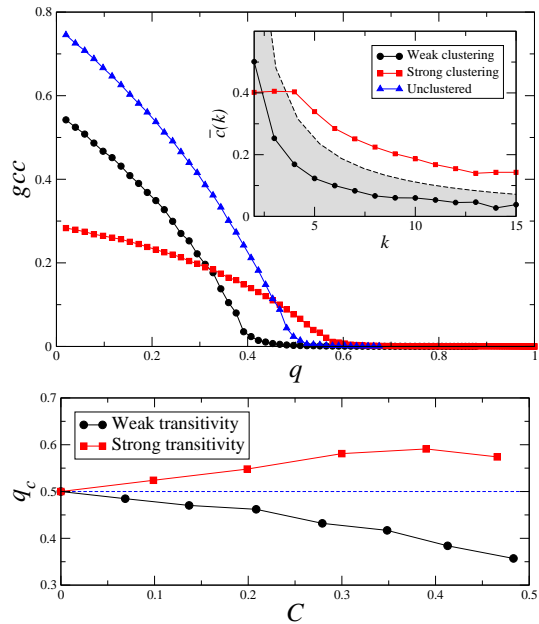

\begin{tabular}{c}
\epsfig{file=Fig3a_paperII.eps,width=7cm}\\
\epsfig{file=Fig3b_paperII.eps,width=7cm}
\end{tabular}
 \caption{Top graph: Bond percolation on networks with exponential
 degree distribution ($\langle k \rangle=2$) and weak clustering (black filled circles),
 strong clustering (red filled squares), and zero clustering (blue filled triangles). For both
 types of clustering the size of the giant component of the original networks is
 smaller than the unclustered one. However, the percolation threshold for the
 network with weak clustering is smaller than the unclustered net whereas it
 is larger for the network with strong clustering, despite the fact that both
 networks have the same value of $\bar{c}$. The inset shows the degree-dependent
 clustering coefficient for both networks. The area depicted in grey indicates the
 limits of weak clustering. Outside this region, the multiplicity of edges is necessarily
 larger than $1$. Bond percolation simulations are performed over a single network of
 size $N=10^5$ and, then, averaged over $50$ different realizations for each value of $q$.
 Bottom graph: Percolation threshold for strong and weak transitivity as a function of the
 global clustering $C=(1-P(1))^{-1}\bar{c}$. The blue dashed line is the percolation threshold
 for the unclustered network ($q_c=1/2$).} \label{fig:6}
\end{figure}
In the strong transitivity regime, the solutions obtained from our
theory become upper bounds to the real size of the giant component
due to the fact that then higher order loops appear. Unfortunately,
it is not easy to extend the analytical calculations to this case
but one can resort to numerical simulations. Fortunately, we can
make use of the algorithm presented in Ref.~\cite{Serrano:2005b},
which allows to generate random networks with a given degree
distribution, a fixed degree-dependent clustering coefficient, and
at the same time, to exert some control on the assortativity level of
the network.

First, it is important to comment on the results by
Newman~\cite{Newman:2003c}, who solved exactly the bond percolation
problem for the one-mode projection of random bipartite graphs. One
of the main results in that study is that clustering, although makes
the giant component smaller, favors its onset, which seems to be
just the opposite result to the one that we obtained in the case of
weak transitivity networks. To solve this puzzle, we first need to
understand how one-mode projection networks are constructed. In a
bipartite network, two types of vertices coexist, for instance
scientists and scientific papers, with connections among them. The
one-mode projection is then constructed by retaining just scientists
and connecting them whenever they coauthor the same paper. That
means that all papers with more than two authors will give place to
cliques of connected scientists. Therefore, edges participating in
triangles will have high multiplicity and the networks so generated
will belong to the strong transitivity class. We shall see in the
following that it is precisely the class the network belongs to
--and not the scalar clustering coefficient-- that determines
whether clustering favors or not the onset of the giant component.

To check this issue, we have generated two networks with identical
degree distribution (exponential with $\langle k \rangle=2$) and
different forms for $\bar{c}(k)$, the first one with weak
transitivity and the second one with strong transitivity. Then, we
study their percolation properties by implementing a bond
percolation experiment. We remove each edge with probability $q$ and
measure the size of the giant component. Fig.~\ref{fig:6} shows the
results of this program. As it is clearly seen, in both types of
networks the size of the giant component is smaller than that of the
unclustered network. However, {\it the percolation threshold for the
weak transitivity class is smaller than the unclustered value,
whereas it is larger in the strong transitivity class, despite the
fact that both networks have the same value for the scalar
clustering coefficient $\bar{c}$}. In Fig.~\ref{fig:6}, we also show the
behavior of the percolation threshold $q_c$ for different values of
the scalar clustering coefficient $C=\bar{c}/(1-P(1))$ (which is
defined in the interval $[0,1]$) for the strong and weak
transitivity cases, which confirms this trend.

\begin{figure}[t]
\epsfig{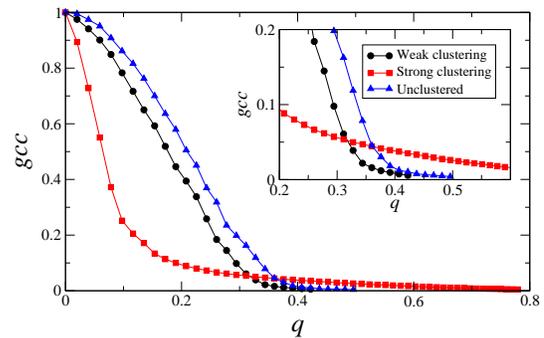}
 \caption{Results of the bond percolation process on networks with a degree distribution
 of the form $P(k)\sim k^{-\gamma}$, $\gamma=3.5$, and strong clustering (red filled squares), weak clustering (black filled circles), and unclustered (blue filled triangles). In all cases, the original networks are single connected components.} \label{fig:7}
\end{figure}
\begin{figure}[t]
\epsfig{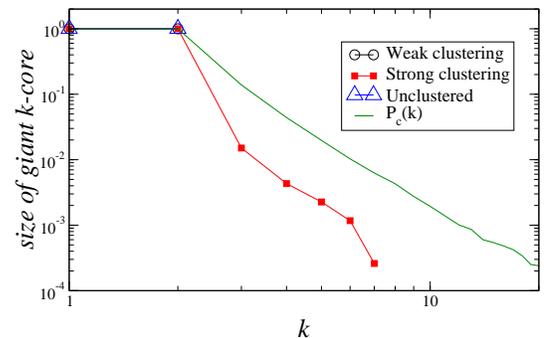}
 \caption{Relative sizes of the giant $k$-cores of networks with
 $P(k)\sim k^{-\gamma}$ and $\gamma=3.5$ for weak clustering,
 strong clustering, and unclustered, respectively.} \label{fig:8}
\end{figure}
Although illuminating, this analysis has the problem that we are
comparing the resilience properties of networks with different giant
components. Despite that the degree distribution is the same for all
the networks, it could not be the case for their giant components,
potentially changing then the location of the critical point. To
overcome this problem, we generate a network with strong
transitivity and extract its giant component. Then, using the degree
distribution of this giant component we generate two more networks,
one without clustering and the other one with weak clustering.
Finally, we check that these two networks are globally connected. We
choose as the degree distribution to generate the strongly clustered
network a function of the form $P(k)\sim k^{-\gamma}$, with
$\gamma=3.5$. This particular form is convenient because using this
distribution the range of degrees is relatively wide and, at the
same time, the network still has a finite percolation threshold. We
also restrict the degrees to values larger than two. Otherwise the
unclustered network would not have a GCC. Fig.~\ref{fig:7} shows the
results for the bond percolation process in these three networks. As
it is clearly seen, the size of the GCC of the highly clustered
network decreases very fast when a small fraction of edges is
removed whereas the unclustered one is more resilient in this
regime. However, as $q$ increases, the unclustered network undergoes
the phase transition (at $q_c=0.4$) whereas the highly clustered net
shows a surprising resilience, decaying very slowly as $q$
increases. On the other hand, the weakly clustered network is always
less resilient than the unclustered one and undergoes the phase
transition at a lower level of damage. These results confirm those
shown in Fig.~\ref{fig:6}

To understand which is the origin of this behavior and the
differences between weak and strong clustering, we need to
understand the kind of structures that are formed depending on the clustering properties. The concept of $k$-core is
particularly suitable for this purpose (see \cite{Dorogovtsev:2006}
for a very nice work on $k$-core percolation in random networks).
The $k$-core of a network is the maximal subgraph such that all its
vertices have $k$ or more connections within the subgraph.
Therefore, $k$-cores are subgraphs which are particularly resilient
to the removal of edges if $k$ is large. In Fig.~\ref{fig:8}, we show
the relative sizes of the giant $k$-cores of the networks of
Fig.~\ref{fig:7} as a function of $k$ compared with the cumulative
degree distribution $P_c(k)=\sum_{k'\ge k} P(k')$. This comparison is
in order because $P_c(k)$ is the maximum possible value the giant
$k$-core can attain. In the cases of weak clustering and unclustered
network, the entire network forms a giant $2$-core but the $k$-cores
for $k>2$ do not exist. This result can be understood using results
from \cite{Dorogovtsev:2006} that state that, in random networks,
the giant $k$-core undergoes a $k$-core percolation transition which is
discontinuous, similar to what happens in first order phase
transitions. In contrast, the strong clustered network has $k$-cores
for $k>2$ which do not vanish, with a constant fraction of $P_c(k)$
belonging to them. This behavior extends up to degree $k=7$, and
the $k$-core finally disappears at $k=8$. This result explains the
strong resilience of the strongly clustered network since, although
small, these $k$-cores with high $k$ are extremely difficult to
break. This suggests that the properties of the critical percolation
threshold in this class of networks should be tied to their $k$-core
percolation properties. This is an issue that deserves further
consideration and will be addressed in a future work.

\section{Percolation in scale-free clustered networks}

\begin{figure}[t]
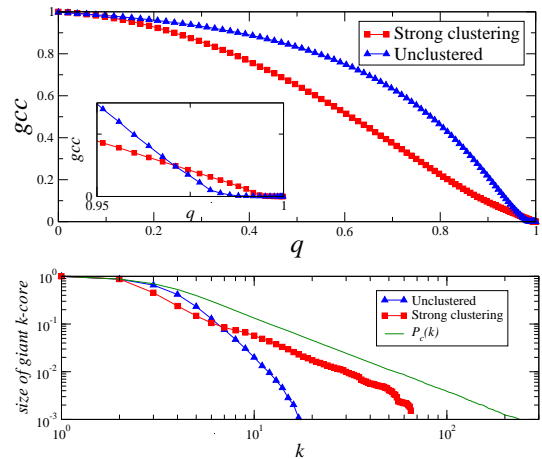

\begin{tabular}{c}
\epsfig{file=Fig6a_paperII.eps,width=7cm}\\
\epsfig{file=Fig6b_paperII.eps,width=7cm}
\end{tabular}
\caption{Top: results of the bond percolation process on a network with a degree distribution
of the form $P(k)\sim k^{-\gamma}$ and $\gamma=2.5$ and its randomized version. The inset shows a detail close to $q=1$. Bottom: relative sizes of the giant $k$-cores of
these two networks. The solid line is the cumulative degree distribution $P_c(k)$.} \label{fig:9}
\end{figure}

Random scale-free (SF) networks with $2<\gamma \le 3$ have the
peculiarity of not having a percolation threshold or, equivalently,
$q_c=1$. This means that one has to remove the totality of the edges
to break the network into disconnected components. In
epidemiological language, it means that any disease, even a low
infectious one, can propagate and infect macroscopic portions of the
population. Given the important implications of this result, it is
necessary to discern whether it can be applied to more general types
of networks. For instance, it has been proved that two-vertices
degree-degree correlations cannot restore a finite epidemic
threshold \cite{Boguna:2003,Vazquez:2003}. However, this result
cannot be applied to networks with high levels of clustering since
the networks used in the demonstration had a vanishing clustering
coefficient in the thermodynamic limit.

As we have already seen in previous sections, weak transitivity
hinders the onset of the giant component, with a condition for its
existence, in the uncorrelated case, given by
\begin{equation}
\frac{\langle k(k-1)\rangle}{\langle k \rangle} >(1+c_0(m_0))
\frac{m_0}{c_0(m_0)} (1-\frac{P(1)}{\langle k \rangle}).
\end{equation}
In the case of SF networks, the term on the left in this inequality
diverges and, therefore, the condition is always fulfilled. In
short, this means that {\it weak transitivity cannot restore a
finite percolation threshold}.

When the network belongs to the strong transitivity class, its
percolation properties are not much different from networks with
bounded fluctuations, except for the fact that the critical
threshold is located at $q_c=1$. In Fig.~\ref{fig:9}, we repeat the
same analysis performed in the previous section but now for a SF
network. We first generate a highly clustered SF network with
$\gamma=2.5$ and extract its giant component. Then, we randomize it
to obtain an unclustered random network preserving the degree
distribution. Finally, a bond percolation process is applied to both
networks. As in the case of $\gamma=3.5$, the clustered network is
less resilient than the unclustered one except for high levels of
damage, where the clustered net is more resilient. In this case, the
fact that we find a finite threshold is due to finite size effects.
Again, the $k$-core analysis reveals a nested structure of $k$-cores
up to $k=65$ following closely the shape of $P_c(k)$. In contrast,
the $k$-cores for the unclustered network decays very fast as $k$
increases. This result implies that {\it strong transitivity cannot
restore a finite percolation threshold either}.

\section{Percolation in real networks: the case of the PGP network}
\begin{figure}[t]
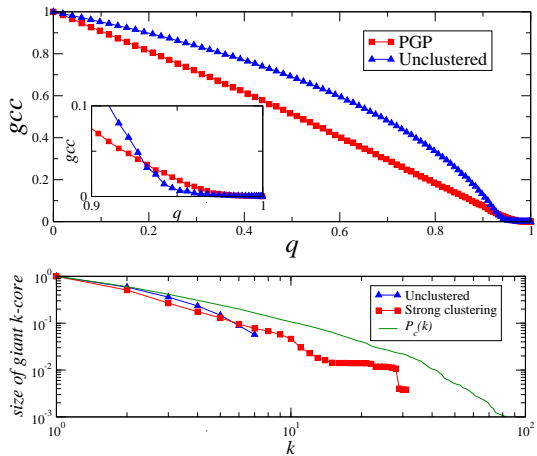

\begin{tabular}{c}
\epsfig{file=Fig7a_paperII.eps,width=7cm}\\
\epsfig{file=Fig7b_paperII.eps,width=7cm}
\end{tabular}
\caption{Top: bond percolation experiment performed on the giant component of the PGP network and on its randomized version. The inset shows a detail close to $q=1$.
Bottom: relative sizes of the giant $k$-cores for the PGP network and its
randomized version, respectively. The solid line is the cumulative degree distribution $P_c(k)$.} \label{fig:11}
\end{figure}

We cannot end this paper without taking a look at the real world and
checking if our results applies also there. To this end, we chose
the Pretty-Good-Privacy (PGP) web of trust analyzed in
Ref.~\cite{Boguna:2004b}. This is a nice example of a large social
network based on trust. It arises as a consequence of the need for
secure communications through the Internet. Without going into great
detail, when a user A wants to send a message to another user B, she
encrypts the message using the public key of user B who, afterwards
uses her private key to decrypt it. In this way privacy is ensured.
However, since everybody can generate his own pair of keys, it is not
possible, in principle, to be sure that the person holding the key
is who she claims to be. An imaginative solution to this problem is
the web of trust. In this web, any user can sign the public key of
another user, meaning that she trusts the other person is who she
claims to be. This procedure generates a publicly available web of trust of users that
have signed the public keys of other users. In principle, this web
is directed. However, since we are interested in social ties, we
filter out those connections that are not reciprocal. In this case,
an edge among two persons is likely to represent a social
relationship between them. After the filtering process,
we obtain a network with $N=57243$ vertices and a giant connected
component of $N_{gcc}=10680$ vertices. The network has a degree
distribution with a heavy tail (although not a pure power law)
extending up to $k\sim 200$ and a degree-dependent clustering
coefficient which is constant up to degree $k=50$ followed by a sharp decay
for $k>50$, with an overall value of $0.5$. This property sets the
PGP network at the heart of the strong transitivity class.

The bond percolation experiment performed on the giant connected component of the PGP network reveals the same type of pattern that we have found before and is
shown in Fig.~\ref{fig:11}. Again, the randomized version is more
resilient except for very high values of $q$, where the PGP net is
more resilient, with the critical point closer to $1$. The $k$-core
decomposition also shows the same type of result. The PGP network
has a nested $k$-core structure extending up to $k=31$, whereas the
randomized network has only $k$-cores up to $k=7$.

We would like to stress that this behavior is by no means exclusive to the PGP network since many networks in the real world belong to the strong transitivity class. The results presented here are extremely relevant in the case of epidemic spreading. On the bad side, they suggest that real clustered networks are more prone to suffer epidemic outbreaks than unclustered networks. Yet, the relative size of the potentially infected population is smaller, which is indeed a positive result. We would also like point out that, the knowledge of the role that the giant $k$-cores have on the percolation properties could be used to design and plan more effective immunization strategies.  

\section{Conclusions}

In this second paper, we have presented an analytical approximation to percolation in weakly clustered networks. Although this formalism is exact only in the limit of weak transitivity, it is an upper bound for the size of the GCC in all cases. Using this approach, we have seen that weakly clustered networks percolate at a higher density of connections as
compared to the unclustered ones. By means of numerical simulations, we
have also proved that the percolation threshold for networks in the
weak transitivity class is smaller than the corresponding to an unclustered network with the same degree distribution. In contrast, this threshold is larger for nets in the strong transitivity class. This means that weak clustering hinders the appearance of the giant connected component whereas it is favored by strong clustering. To understand which is the origin of this behavior and the differences between weak and strong clustering, we have explored the structural organization of networks through their $k$-core decomposition, finding important differences among the two classes. In the case of scale-free networks, we have seen that neither weak nor strong transitivity can restore a finite percolation threshold.
We have checked our results using a real social network, finding a very good agreement.

To summarize, in this paper and the preceding one we have developed
a full theoretical approach to clustering in complex networks.
We hope that these developments will improve the topological
characterization of complex networks but also that they will help to produce
a better understanding and a more realistic modelling of the dynamical
processes that conform or use them.

\begin{acknowledgments}
This work has been
partially supported by DGES, Grant No. FIS2004-05923-CO2-02 and
Generalitat de Catalunya Grant No. SGR00889. M. B. thanks the School
of Informatics at Indiana University, where part of this work was
developed.
\end{acknowledgments}


\end{document}